\documentclass[doublecol]{epl2}
\usepackage{graphicx,amsmath,amssymb}

\title{Transmission thresholds in time-periodically driven nonlinear 
disordered systems}
\shorttitle{Transmission thresholds in driven nonlinear disordered systems} 

\author{Magnus Johansson\inst{1,5} \and Georgios Kopidakis\inst{2,5}
\and Stefano Lepri\inst{3,5} \and Serge Aubry\inst{4,5}}
\shortauthor{M. Johansson \etal}

\institute{
  \inst{1} Department of Physics, Chemistry and Biology (IFM), Link\"{o}ping 
University, SE-581 83 Link\"{o}ping, Sweden\\
  \inst{2} Department of Materials Science and Technology, University of 
Crete, GR-71003 Heraklion, Greece\\
  \inst{3} Istituto dei Sistemi Complessi, Consiglio Nazionale delle 
Ricerche, via Madonna del piano 10, I-50019 Sesto Fiorentino, Italy\\
  \inst{4} Laboratoire L\'eon Brillouin, 
CEA Saclay, 91191 Gif-sur-Yvette, France\\
  \inst{5} Max Planck Institute for the Physics of Complex Systems, 
N\"{o}thnitzer Str. 38, D-01187 Dresden, Germany
}

\abstract{
We study energy propagation in locally time-periodically driven disordered 
nonlinear chains. For frequencies inside the band of linear Anderson modes, 
three different regimes are observed with increasing driver amplitude: 
1) Below threshold, localized quasiperiodic oscillations and no spreading; 
2) Three different regimes in time close to threshold, with almost regular 
oscillations initially, weak chaos and slow spreading for intermediate times, 
and finally strong diffusion; 3) Immediate spreading for strong 
driving. The thresholds are due to simple  bifurcations, obtained 
analytically for a single oscillator, and  numerically as turning-points of 
the nonlinear response manifold for a full chain. Generically, the threshold 
is nonzero also for infinite chains.}    

\begin{document}

\maketitle

Energy transport is rather well understood in linear discrete Hamiltonan 
systems, where general solutions 
are linear combinations of eigenmodes, and the transport properties are 
completely determined by the nature of the spectrum of 
eigenfrequencies and the corresponding 
eigenmodes. For 
spatially periodic systems the 
linear spectrum is absolutely continuous and the 
eigenmodes are planewaves, implying that
any initially localized wavepacket will generally 
disperse throughout the system, with amplitude vanishing to zero 
at infinite 
time. Conversely, when the linear system is strongly disordered we have 
Anderson localization, characterized by a purely discrete linear 
spectrum with square summable (localized) eigenmodes. 
Then, initially localized wavepackets
do not spread to zero and there is no energy diffusion. 

The aim of this paper is to understand how some aspects of 
the transport properties of systems with linear Anderson localization are 
affected when nonlinearity is added. 
Obviously, for a nonlinear system, general 
solutions are not just linear combinations of eigenmodes. Still 
however, 
special (time-periodic) solutions generally exist, 
which continue single linear eigenmodes from the small-amplitude limit, and 
thus in some sense correspond to nonlinear eigenmodes. Due to the 
nonlinearities, the 
frequency of such a continued solution varies with amplitude, and 
each time it crosses a resonant linear eigenfrequency a 
new peak appears 
at the location of the corresponding linear mode. Since there are 
infinitely many resonant frequencies 
in any finite frequency interval, the continued solution develops infinitely
many new peaks, and thus  
the strict continuation of a localized Anderson mode 
becomes a {\em spatially extended} mode 
for any non-zero amplitude \cite{KA99}. 
Experimentally, such {\em nonlinear delocalization} was reported in 
\cite{Pertsch} for a two-dimensional fiber array. At larger amplitude, the 
spatial density of the peaks increases so that even in-between peaks 
the amplitude is never small, thus allowing the transportation of a 
substantial energy current by phase torsion \cite{Aubry97}. However, despite 
the existence of these extended solutions it has been 
shown \cite{Albanese,KAPRL,KA00}, that spatially {\em localized} time 
periodic solutions also persist with nonlinearities, but when the frequency 
is inside the linear band they are not in strict continuation of the linear 
modes. Instead, the frequency of a localized solution becomes a 
discontinuous function of its amplitude, 
with a gap located 
near each linear eigenfrequency. Although there are infinitely many gaps 
since the eigenfrequencies form a dense 
set in the spectrum, the gap widths 
become exponentially small as functions of the spatial distance of the 
resonance location, and as a consequence the 
``allowed'' frequencies belong to a (fat) Cantor set with nonzero measure
\cite{Albanese}. 
Indications of experimental excitation of 
a sequence of such ``{\it intraband discrete breathers}'' were given in 
\cite{Pertsch}. Finally, when the nonlinearity is strong enough to drive the 
frequency outside the linear band, ``extraband discrete breathers'' 
(discrete solitons) are formed, and localization is generally 
{\it enhanced} as has been seen experimentally for several types of 
photonic lattices \cite{Pertsch,Segev,Lahini}

Thus, this
indicates that nonlinearities on one hand maintain the existence of
localized solutions which could trap energy forever, 
but on the other hand also generate new extended solutions 
which could transport energy. To investigate the possible energy 
diffusion, numerical experiments have studied the time evolution 
of an initially localized wave packet \cite{Shep93,Molina,PS08,KKFA,GMS,FKS}. 
This may be analyzed by expanding the 
solutions in the basis of Anderson modes, which then become 
coupled by nonlinear terms. It was early conjectured \cite{Shep93} 
that this coupling, if strong enough, would allow the 
energy to diffuse from one Anderson mode to another, leading to a (subdiffuse) 
spread throughout the system. Several 
numerical studies apparently 
confirmed that
the second moment of the energy distribution of a single wave packet 
may diverge as a function of 
time \cite{Molina,PS08}. 
However, this  does {\em not} necessarily imply that
its amplitude vanishes at infinite time. 
By observing 
the time evolution of the participation number, which measures the 
localization 
length of the core of the wavepacket, it was shown \cite{KKFA} that for 
large enough initial amplitude the 
participation number remains finite,  although the second moment 
may diverge if some part of the energy spreads to infinity. The 
observations in \cite{KKFA} also suggested that the wavepacket
may have a 
limit profile (possibly with infinite second moment) which is an almost 
periodic solution (with a discrete Fourier 
transform) and 
zero Lyapunov coefficients analogous to KAM tori in finite 
systems. This conjecture is supported by rigorous 
results proving the 
existence of such solutions in random DNLS systems \cite{spencer,BW08}. 

This paper is devoted to another test for the energy transport property: 
the transmission 
through a system with a {\em time periodic driving force} 
locally applied. 
While the above arguments referred  to finite energy 
wavepackets in systems at zero temperature, the external 
driving force now acts as an energy source.
We have investigated 
several nonlinear models with linear Anderson localization, but
focus on two prototype examples: the Discrete Nonlinear 
Schr\"odinger (DNLS) and Fermi-Pasta-Ulam (FPU) models, in both cases  
with driving at a single edge site of a semi-infinite chain. 
 {Numerically, we 
choose the chain length $N$ to be sufficiently large to have no 
observable change  when increasing $N$.} 
 {Since we aim at describing a qualitative scenario generically valid 
for any typical disordered chain, we do not average over 
different realizations.}

A DNLS chain with random on-site potential and
 non-parametric periodic (harmonic) driving at a boundary site is described 
by the dynamical equations
\begin{equation}
\mathrm{i} \dot{\psi}_{n} - V_n \psi_n +C(\psi_{n+1}+\psi_{n-1})+|\psi_n|^2 \psi_n 
= A_n \mathrm{e}^{\mathrm{i} \omega t}, 
\label{DNLS1}
\end{equation}
where $C$ is the coupling constant, and the amplitude $A_n=A \delta_{n0}$ of 
the driving force at frequency $\omega$ is non-zero only at site $n=0$. 
We take $V_n$ to be random variables with uniform 
distribution in the interval $[-V, +V]$. This 
gives a linear spectrum (for undriven chain) between $[-2C-V, +2C+V]$. 
The explicit time-dependence of the force in eq.\ (\ref{DNLS1}) is removed
by transforming to a frame rotating with frequency 
$\omega$,  $\psi_n(t) = \phi_n(t) \mathrm{e} ^{\mathrm{i} \omega t}$, yielding 
\begin{equation}
\mathrm{i} \dot{\phi}_{n}-(\omega+ V_n) \phi_n+C(\phi_{n+1}+\phi_{n-1})+|\phi_n|^2 
\phi_n = A_n .
\label{DNLS2}
\end{equation}
For a given initial condition $\phi_n(0)$, there are three independent 
parameters: $A/C$ (driving strength), $V/C$ (disorder strength), $\omega/C$ 
(driving frequency). (We may always put $C=1$ by rescaling time and 
wavefunction.) Equation (\ref{DNLS2}) can be derived from the Hamiltonian  
\begin{eqnarray}
\mathcal{H} =\sum_{n=0}^{+ \infty} \left\{(\omega+V_n)|\phi_n|^2
- C(\phi_{n+1}\phi_{n}^*
+ \phi_{n+1}^*\phi_{n}) 
\right. \nonumber \\ 
\left.
-\frac{1}{2} |\phi_{n}|^4+
A_n \phi_{n}^{\ast }+ A_n^* \phi_{n}
 \right\}, 
\label{HNS}
\end{eqnarray}
since 
$\mathrm{i}\dot{\phi}_n = \frac{\partial {\mathcal H}}{\partial \phi_n^\ast}$, 
$\mathrm{i}\dot{\phi}^\ast_n = -\frac{\partial {\mathcal H}}{\partial \phi_n}$.
Thus, even in presence of driving 
${\mathcal H}$
is a conserved quantity, 
characterizing the initial condition. If we choose ${\phi_n}(0)\equiv 0$ and 
apply the driving instantaneously at $t=0$, then $\mathcal{H}=0$. 
However, if we 
slowly increase the force from zero to its final value $A$ during a transient 
time, then $\mathcal{H}$ is not constant during the transient, and 
generally
$\mathcal{H}\neq 0$ after the transient. We measure the "amount of 
excitation" in the 
system at a given time from the total norm (or power in optics) 
\begin{equation}
{\mathcal N}(t)=\sum_{n} |\phi_n|^2, 
\label{norm}
\end{equation}
which is a conserved quantity only in absence of driving. 

If all parameters  
$|A/C|$, $|V_0/C|$, $|\omega/C|$ are large, we can neglect the
coupling to the rest of the lattice and only consider the 
{\it single anharmonic oscillator} at $n=0$:
\begin{equation}
\mathrm{i} \dot{\phi}_{0} - (\omega+ V_0) \phi_0 +|\phi_0|^2 \phi_0 
= A .
\label{single}
\end{equation}
The nature of stationary 
solutions (constant $|\phi_0|$) to eq.\ (\ref{single}) is wellknown (see, 
{\it e.g.}, ref.\ \cite{Rigo}). Due 
to conservation of $\mathcal{H}$, it is integrable. Expressing 
$\phi_0 = r_0 \mathrm{e}^{\mathrm{i} \theta_0}$ ($r_0 \geq 0$ and 
$\theta_0$ real) yields 
$\dot{r}_0 = -A \sin {\theta_0}$; 
$\dot{\theta}_0 + \omega+V_0-r_0^2 = -\left(A/r_0\right)  \cos \theta_0$. 
Thus, a stationary solution ($\dot{r}_0 =0$) 
in presence of driving ($A\neq 0$) can only exist if $\theta_0 =0$ or $\pi$, 
corresponding to oscillations inphased resp.\ antiphased to the driving force. 
They are obtained by solving the cubic equation 
$r_0^3 - (\omega+V_0) r_0 \mp A =0$, where upper (lower) sign corresponds 
to $\theta=0$ ($\pi$). 
When $A=0$, 
the solutions are $r_0=0$ ("linear solution", $\mathcal{H}=0$) and 
$r_0=\sqrt{\omega+V_0}$ if $\omega+V_0 >0$ ("nonlinear solution", 
$\mathcal{H}= (\omega+V_0)^2 / 2$). For $A>0$, the cubic equation for 
$\theta = 0$ always has one (stable) 
solution for $r_0 \geq 0$, which continues the 
"nonlinear solution" towards larger $r_0$  
when  $\omega+V_0 >0$ and the "linear solution" 
when $\omega+V_0 < 0$. In addition, for small $A>0$ and  $\omega+V_0 >0$
the cubic equation for $\theta=\pi$ 
has two solutions with $r_0>0$, 
one stable continuing the "linear solution" towards 
 larger $r_0$ and one
unstable continuing the "nonlinear solution" towards smaller $r_0$. However, 
at a critical driving strength 
$A=A^{(c)} = 2(\omega+V_0)^{3/2}/\left( 3\sqrt {3}\right)$ these solutions 
bifurcate, so that for $A>A^{(c)}$ only the "nonlinear solution" with 
$\theta = 0$ remains. 

The significance of this bifurcation is, that if 
$A$ is increased slowly (adiabatically) from zero for a zero initial
condition when $\omega+V_0 >0$, the solution will follow the stable solution 
with $\theta=\pi$ until it disappears at $A=A^{(c)}$. Above this 
{\em threshold} the 
solution cannot jump dynamically to the other stable stationary 
solution with $\theta = 0$
since it has larger ${\mathcal H}$, but instead $r_0$ becomes periodic with a 
new 
oscillation frequency (corresponding to a quasiperiodic solution in the 
non-rotating frame). By contrast, for $\omega+V_0 < 0$ there is always 
only one solution which continues the "linear solution" for all $A$. 

For the full chain, an analogous threshold can be described by
the nonlinear response manifold (NLRM) technique, 
used in \cite{KAPRL,KA00} to calculate time-periodic 
breathers for a non-driven disordered lattice, and in \cite{MKA06} to 
analyze transmission thresholds in driven nonrandom Klein-Gordon 
chains \cite{leon}. 
For the 1D DNLS model, it can be simply implemented numerically.
Looking for real ({\it i.e.}, carrying no current) 
stationary solutions with $\dot{\phi_n}=0$, eq.\ (\ref{DNLS2}) 
can be written 
as a 2D map with $x_n = \phi_n$, $y_n=\phi_{n+1}$, which we may iterate 
backwards from $n=N$ to $n=1$,
$$
\left(x_{n-1}, y_{n-1}\right) = 
\left(\frac{(\omega+V_n-x_n^2)}{C}x_n - y_n, x_n \right) .
$$ 
Using as map  initial (boundary) condition $(x_N, y_N)=(0,\epsilon)$ for a
number of  different $\epsilon$ (typically $\sim 1000$ with max 
$\epsilon$ $\sim 10^{-12}$ for $N\simeq 50-100$
sites)  gives  a number of points belonging to a  manifold at the end of
iteration at $(\phi_0, \phi_1)$. Combining this with  the equation at the edge
site $n=0$,   $A = -(\omega+V_0-x_0^2)x_0 + C y_0$, and plotting, {\it e.g.},
$x_0=\phi_0$ as a  function of $A$ gives a projection of the NLRM, from which 
we
may obtain {\em  all real stationary solutions at given driving strength $A$} 
as
intersections  of the NLRM  with the vertical line at $A$. 
(The above set  of map initial conditions yields only half
the NLRM, the {other half} is obtained by adding 
$\phi_0(-A)= -\phi_0(A)$  {but in order to not overload the figures 
it is 
not shown below})  As discussed in \cite{MKA06}, 
{\em each turning  point (TP)} of the NLRM can be associated with a
threshold-like behaviour.  
{Increasing the system size may yield more loops in the NLRM, 
but the structure of the first loops, and in particular the first TP, are 
generally found to remain unchanged for $N \gtrsim 100$.}

We now illustrate numerically, how
these NLRM TPs are related to {\em transmission thresholds} 
in the dynamics. Attempting to adiabatically follow the continuation 
of the linear stationary ({\it i.e.}, with same frequency as driving force) 
solution 
as long as it exists, we increase the driving strength slowly from 
zero to its final value. as 
$A(t)=A(1-\mathrm{e}^{-t/\tau})$, with typically $\tau \sim 100-1000$. 
We choose a rather strong disorder, $V=2.5$, in order to have 
well localized linear modes (we put $C=1$). 
To be concrete, we show results obtained for 
one particular realization of the disorder, for which the on-site potential 
of the driving site is $V_0 \approx 2.48977$. 
\begin{figure}
\includegraphics[height=0.402\textwidth,angle=270,clip]{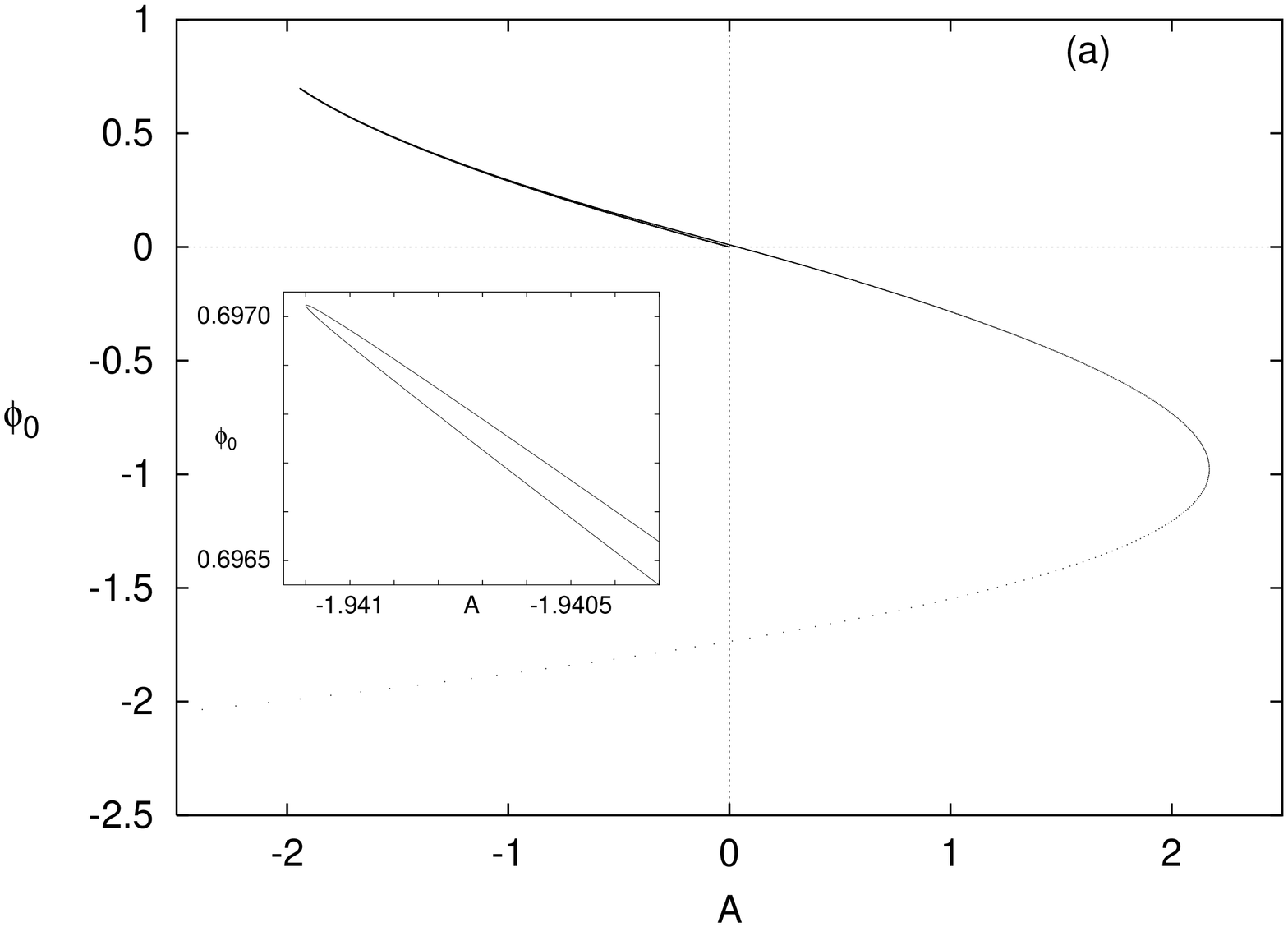}
\includegraphics[height=0.402\textwidth,angle=270]{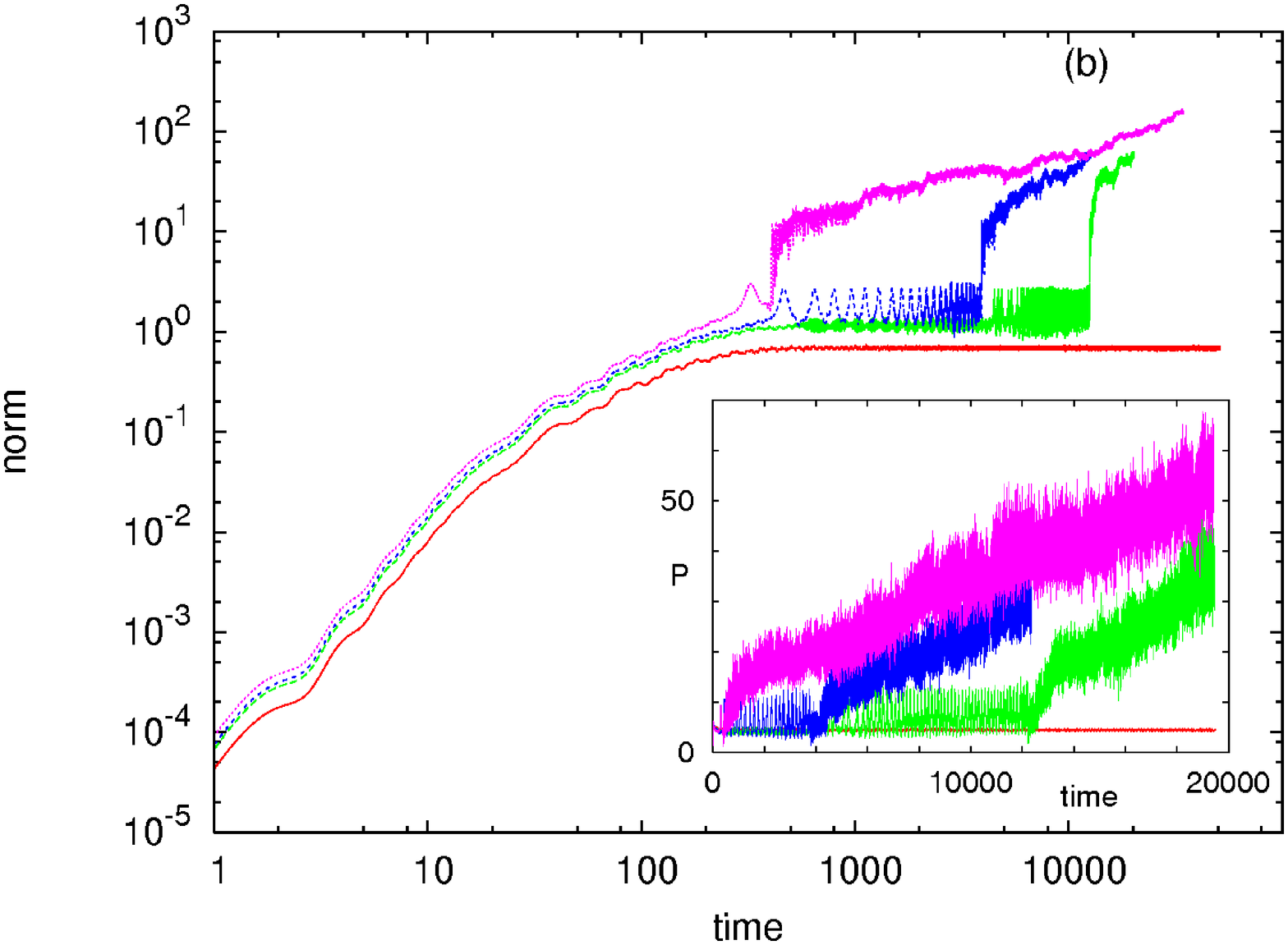}
\includegraphics[height=0.212\textwidth,angle=270]{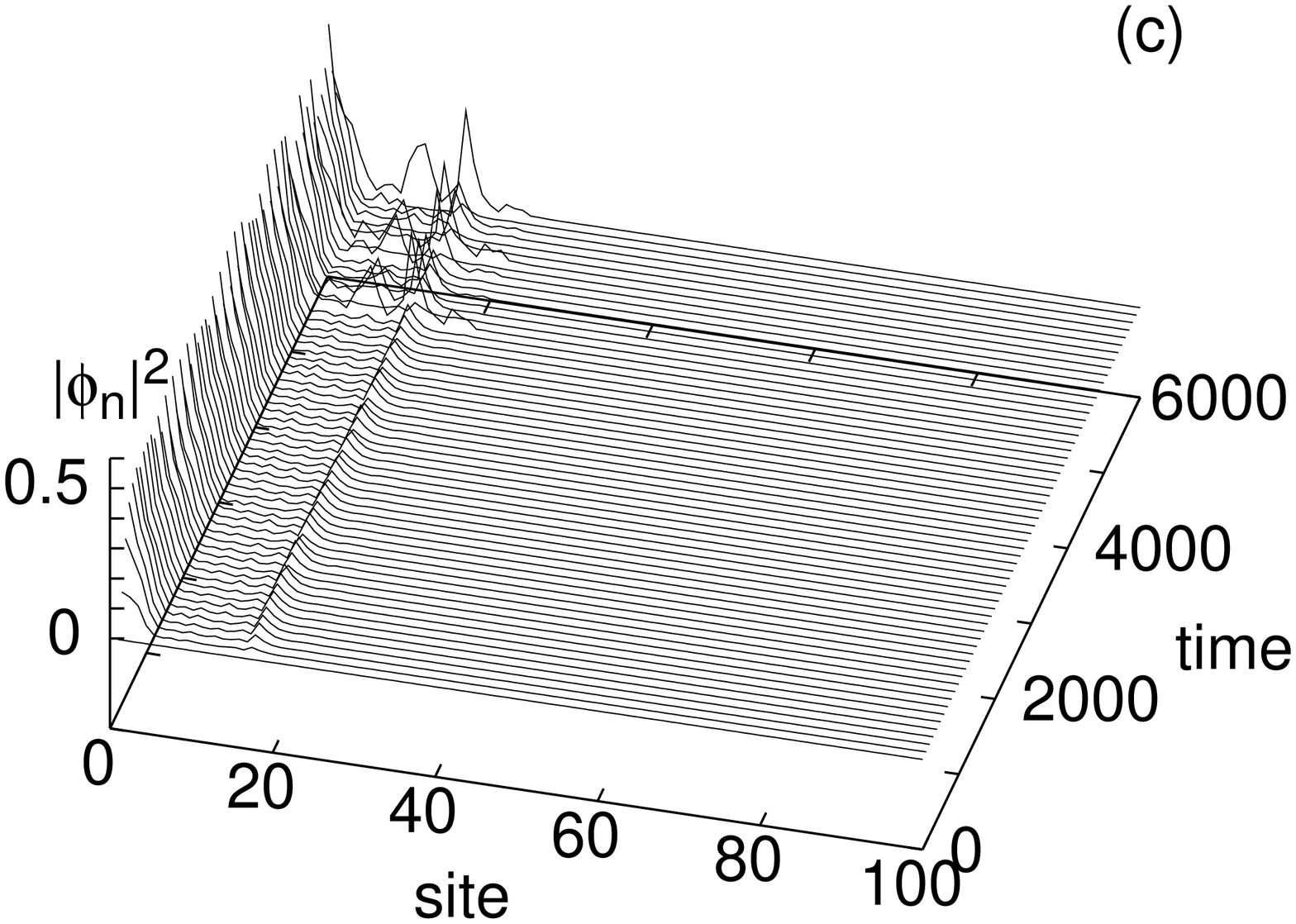}
\includegraphics[height=0.212\textwidth,angle=270]{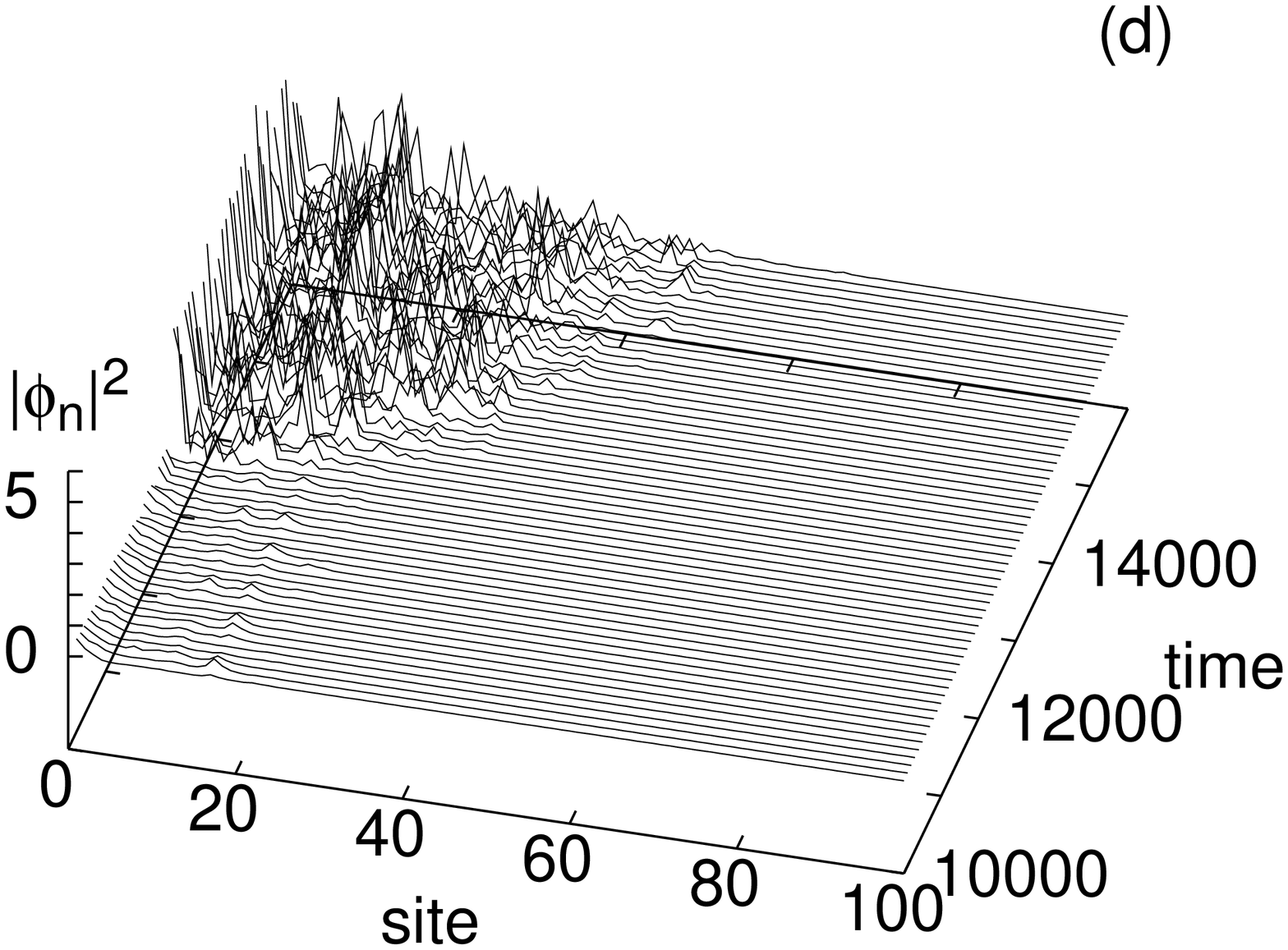}
  \caption{(a) {(Half)} NLRM for a random DNLS chain with $V=2.5$ 
and $\omega=0$. 
The apparent line through 
the origin is in fact a very thin loop, continuing
the linear solution from the origin towards negative $A$ 
until 
it reaches the first TP at $(-1.9411, 0.697025)$ 
(inset in (a)). 
It then 
returns to the right, crosses $A=0$ at a small positive $\phi_0$,  
reaches a second TP at (2.17, -0.97), and then escapes towards 
large negative values (not shown).  
(b) Time evolution of norm (main figure) and participation 
number (inset) for, from bottom to top, $A=1.5$, 
$A=1.9$ (just below first TP), $A=2.0$ (between 
first and second TP), and $A=2.2$ (above second TP). Transient time 
$\tau=100$. 
{(c), (d)}: Shape of the solution for $A=1.9$, showing 
(c) the first transition from almost regular to slowly spreading, 
and (d) the second transition to strongly diffusive, behaviour. 
}
\label{fig5}
\end{figure}
In fig.\ \ref{fig5} we show (half) the NLRM and time-evolution of total 
norm (\ref{norm}) and participation number 
$P={\mathcal N}^2/\sum |\phi_n|^4$
for different $A$ for a driving frequency in the 
middle of the linear band, $\omega=0$.

When the driving strength is far 
below the first TP of the NLRM ($A=1.5$ in fig.\ \ref{fig5}(b)), 
the response is essentially linear-like and, 
if the force is increased sufficiently slowly, the solution is essentially 
a stationary solution at the driving frequency, localized close 
to the driving site and belonging to the stable
(lower) branch of the first (left) thin bifurcation loop in  
fig.\ \ref{fig5}(a)). 
The small 
(quasiperiodic) oscillations are due to non-adiabaticity, and will decrease 
if $\tau$ is increased. Thus, this represents the 
sub-threshold-type of non-transmitting dynamics. 
 {We have confirmed that this regular behaviour remains also for 
considerably longer integration times,  {\it e.g.}, $10^6$.}

For $A$ just below the first TP (fig.\ \ref{fig5}(c), (d)), 
we observe a critical behaviour, where the small oscillations 
due to non-adiabaticity 
generally are weakly chaotic 
and cause the orbit to
slowly diffuse in phase-space. At $t\sim 5000$ it jumps to a different 
{phase-space region}, with bounded but larger oscillations 
(fig.\ \ref{fig5}(c)). We associate this 
with the existence of a neighboring stable stationary solution on
the second loop of the NLRM (these solutions are close due to the thinness 
of the first loop). However, finally
($t\sim 13000$, fig.\ \ref{fig5}(d)), the 
solution escapes into a continuously spreading, strongly 
chaotic 
state with rapid increase of ${\mathcal N}$ and $P$. 
A similar scenario is observed also for $A$ between the first and second TPs 
($A=2.0$ in fig.\ \ref{fig5}(b)), 
although in this case the weakly chaotic 
oscillations always develop as soon as $A$ passes 
the first TP, even for larger $\tau$.

Finally, for $A$ above the second TP ($A=2.2$ in fig.\ \ref{fig5}(b)), the
spreading starts essentially immediately, and relevant 
measures like norm and participation number 
generally grow as $t^\alpha$, where $\alpha$ is close to 0.5 indicating a
diffusive-like behaviour 
(for some different cases we obtained exponents 
between approximately 0.44 and 0.60). 
  
\begin{figure}
\includegraphics[height=0.431\textwidth,angle=270]{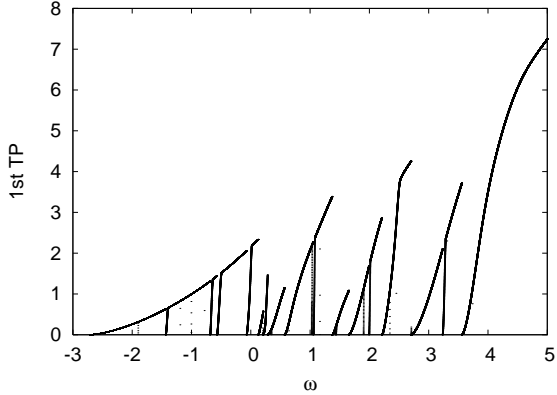}
  \caption{Adiabatic threshold  $A_{\mathrm{th}}(\omega)$, determined from the 
first TP 
of the NLRM, for the same DNLS chain as in Fig\ \ref{fig5}. The resolution in 
$\omega$ is $10^{-4}$ and chain length $N=100$. 
}
\label{figTP}
\end{figure}
The main qualitative features discussed above are generic, although in 
particular the quantitative behaviour in the critical regime depends on the 
detailed structure of the NLRM, which for other parameter values 
may be considerably more complicated 
than in fig.\ \ref{fig5}(a). But generally, for 
each random chain (fixed disorder realization and $V$) we can identify 
an adiabatic transmission threshold $A_{\mathrm{th}}(\omega)$ from the first TP 
of the 
NLRM, remembering 
that the real observed threshold is somewhat 
smaller due to 
nonadiabatic effects (finite $\tau$). The 
dependence  $A_{\mathrm{th}}(\omega)$  exhibits sharp 
variations, as shown in fig.\ \ref{figTP} 
for the chain used in fig.\ \ref{fig5}. For an 
infinite chain, $A_{\mathrm{th}}(\omega)$  is conjectured to be an upper 
semicontinuous function 
(larger than or equal to its limit at any point), which should become zero at 
any frequency resonant with a 
linear mode ({\it i.e.}, on a dense set of points), but nonzero for most
frequencies and in average. In fig.\ \ref{figTP} we can clearly observe 
the stronger resonances corresponding to modes located close to the driving 
site, 
while weaker resonances from distant modes should 
cause similar effects more difficult to observe numerically. We also confirmed 
that {the transmission threshold decreases towards zero} when 
approaching the lower band edge, as indicated by fig.\ \ref{figTP}. However, 
for frequencies smaller than the smallest resonance frequency (lower band 
edge for an infinite system, and $\omega \approx -2.7$ for the finite system 
in fig.\ \ref{figTP}), there is no TP at all of the NLRM, and the numerical 
simulations show complete absence of transmission even for very large 
driving forces. 
 
To illustrate that the above scenario is indeed generic, and not due to 
special DNLS properties like conservation of ${\mathcal H}$, 
we now discuss a random FPU chain with $N$ sites and a cubic
nonlinear force. The equations of motion are
\begin{eqnarray}
\ddot u_n =K_n(u_{n+1}-u_n) - K_{n-1}(u_n-u_{n-1}) 
\nonumber \\
+(u_{n+1}-u_n)^3-(u_{n-1}-u_n)^3, 
\label{fpu}
\end{eqnarray}
where $u_n$ is the displacement of  site $n$
and
$K_n$ 
are random coupling constants, 
independent and uniformly 
distributed in $[K_{\mathrm{min}},K_{\mathrm{max}}]$. In the following we fix 
$K_{\mathrm{min}}=1$ and $K_{\mathrm{max}}=4$ (we put $K_0=1$).
To simulate 
an impinging wave we impose on one end 
the boundary condition
$
u_0(t) = A \cos \omega t 
$, 
while free boundary conditions, $u_{N+1}=u_N$, 
are enforced on the other side of the chain.

An important
difference between the FPU and DNLS models is that, with free
boundary conditions on both sides,   
the FPU linear eigenmodes are well localized only for large enough 
frequencies, 
since the localization length $\xi$ 
diverges as $\xi \sim \omega^{-2}$ in the limit  
$\omega\to 0$ \cite{MI70}.
More precisely, in a chain of $N$ sites, all eigenvectors for 
which  \cite{MI70}
\begin{equation}    
  \omega \; < \; \omega_{\mathrm L}  \;\equiv\;
  \sqrt{\frac{8 \langle K^{-1} \rangle}
  {N(\langle K^{-2} \rangle-\langle K^{-1}\rangle^2) } }
\label{omegal}
\end{equation}
are in practice extended (phonon) modes. 
Therefore, we consider
driving frequencies in the upper part of the linear spectrum, 
$\omega_{\mathrm L} < \omega < \omega_{\mathrm{max}}$ where 
$\omega_{\mathrm{max}}$ is the band-edge. 

\begin{figure}
\begin{center}\leavevmode
\includegraphics[width=0.842\linewidth,clip]{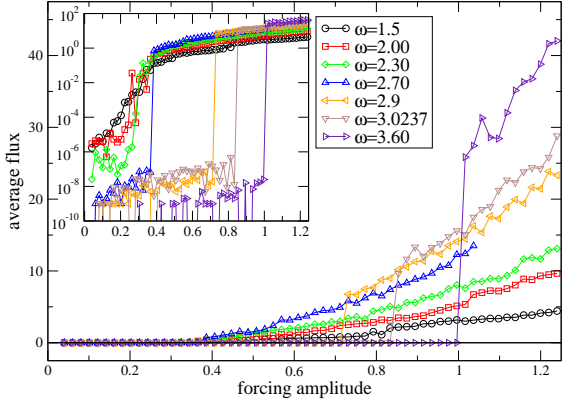}
\end{center}
\caption{Average fluxes  $\overline{j}$ 
versus amplitude driving $A$ for different 
frequencies $\omega$. $N=512$, $N_D=50$,  $\gamma=0.1$. 
The averages are performed on
about $2\cdot  10^4$ periods. Inset: same data in lin-log scale. 
Notice that the jump can be six orders of
magnitude.
}
\label{flx}
\end{figure}

In a 
series of numerical experiments we initialized the chain at rest 
($u_n \equiv \dot u_n \equiv 0$) and 
increased smoothly the driving amplitude from 0
to 
$A$ with a constant rate,  
$
u_0 = A\cos(\omega t)\left[1-\mathrm{e}^{-t/\tau}\right] 
$, 
with typically $\tau=10$. 
To simulate a semi-infinite lattice and 
observe a stationary state in the transmission regime,  we 
steadily removed the energy injected from the driver
by adding a damping term $-\gamma \dot u_n$ to eq.\ (\ref{fpu})
for a number $N_D$ of rightmost sites  ($\sim 10\%$ of the
total). 
As indicator, we used the local energy  flux
\begin{equation}
j_n=\frac12 (\dot u_n+\dot u_{n+1})
\left[K_n(u_{n+1}-u_n)+(u_{n+1}-u_{n})^3\right]. 
\label{flux}
\end{equation} 
After a transient
of $\sim 1000$ driver periods, 
the site- and time-averaged 
energy flux 
$
\overline{j} = \left(\sum_{n=1} ^{N-N_D} \overline{j_n}\right)/\left(N-N_D\right) 
$ 
was computed (the bar denotes time average).
Figure \ref{flx} shows 
results for several values of $\omega$ 
($\omega
\gg \omega_{\mathrm L} \sim 0.44$), for a fixed disorder realization. 
As in
the DNLS case, a well-defined transmission
threshold $A_{\mathrm{th}}$ exists for every frequency.  
Its exact value depends on the specific 
disorder realization (becoming small if there is a resonant mode 
close to the edge), but 
the qualitative behaviour is the same. 
Moreover, $A_{\mathrm{th}}$ is insensitive to variations of the size $N$ 
{(we checked  sizes between $N=512$ and $N=4096$)} 
and
dissipation at damped sites  $N_D$.
Similarly to the DNLS case (cf.\ fig.\ \ref{figTP} 
{and the analytical 
result $A^{(c)}$ for a single oscillator}), there is an average 
tendency (disregarding the individual resonances) 
for  $A_{\mathrm{th}}$ to increase with increasing frequency; note however 
that in the transmitting regime, the flux for a given 
driver amplitude {\em increases} with $\omega$.

Also for FPU the 
threshold is related to a TP of the NLRM, which may be
easily computed in a rotating-wave approximation
(a more computationally expensive exact calculation could be done 
analogously to
refs.\ \cite{KAPRL,KA00,MKA06}). 
Looking for 
solutions of the form 
$u_n(t)=U_n \cos\omega t$ 
and approximating $\cos^3 \omega t \approx (3/4) \cos\omega t$ in 
eq.\ (\ref{fpu}), we get 
\begin{eqnarray}
-\omega^2 U_n =
K_n(U_{n+1}-U_n) - K_{n-1}(U_n-U_{n-1})
\nonumber\\
+\frac{3}{4}(U_{n+1}-U_n)^3-\frac{3}{4}(U_{n-1}-U_n)^3. 
\end{eqnarray}
This equation can be solved with respect to $U_n-U_{n-1}$ 
yielding a two dimensional backward map 
$(U_{n-1}, V_{n-1}) = (U_n - F(V_n,U_n),  U_n)$, 
where the function $F$ is 
obtained by solving 
a cubic equation. 
\begin{figure}[ht]
\begin{center}\leavevmode
\includegraphics[width=0.83\linewidth,clip]{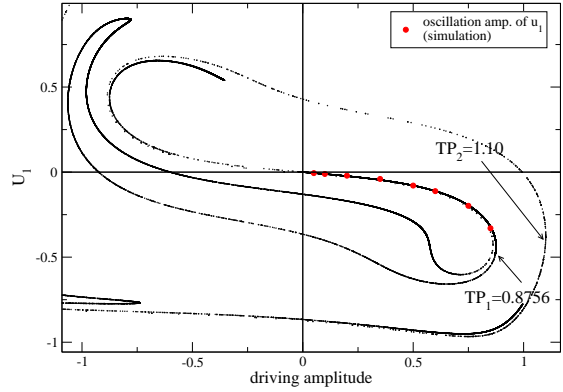}
\end{center}
\caption{Projection of the approximate FPU NLRM, obtained by iterating a 
few hundred points in an interval of size $\epsilon \sim 10^{-8}$ around the 
origin, for
$\omega= 3.0237$ and the same disorder realization as in 
fig.\ \ref{flx}. Large points show
the maximal oscillation amplitudes of the first particle, 
measured from simulation.
}
\label{manfpu}
\end{figure}
In fig.\ \ref{manfpu} we show the last iterate $(U_0,V_0)$
(note that $U_0=A$) of a set of trajectories 
started along the unstable manifold of the origin. 
The curve  is locally linear around $A=0$ (linear response) and 
then bends and turns wildly similar as discussed above.  
This computed manifold agrees very well with
the data obtained from simulation when the driving is switched on very 
slowly (here 
$\tau=400$).
As can be seen, the first TP ($TP_1$ in fig.\ \ref{manfpu}) 
is at $A = 0.8756$ which is in 
excellent agreement with the observed transmission threshold 0.878.

\begin{figure}[ht]
\begin{center}\leavevmode
\includegraphics[width=0.81\linewidth,clip]{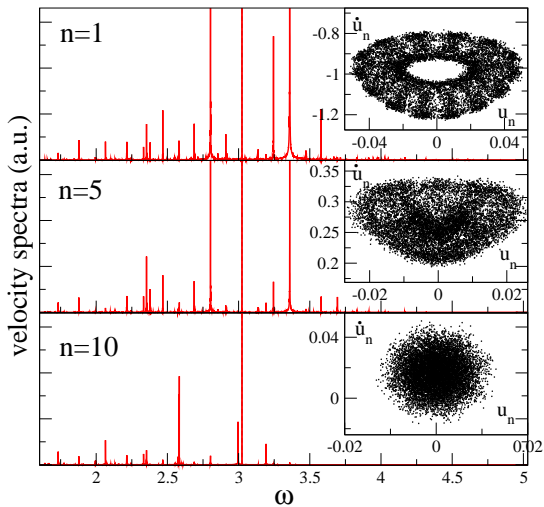}
\end{center}
\caption{
Fourier power spectra of the velocities $\dot u_n$ at sites $n=1,5,10$ 
for $\omega=3.0237$ and  $A=0.850$ (just below threshold). 
The insets
show the Poincar\'e sections $(u_n(kT),\dot u_n(kT))$ ($k$ integer)
where $T=2\pi/\omega$ is the driving period. 
}
\label{qperio2}
\end{figure}

\begin{figure}[ht]
\begin{center}\leavevmode
\includegraphics[width=0.74\linewidth,clip]{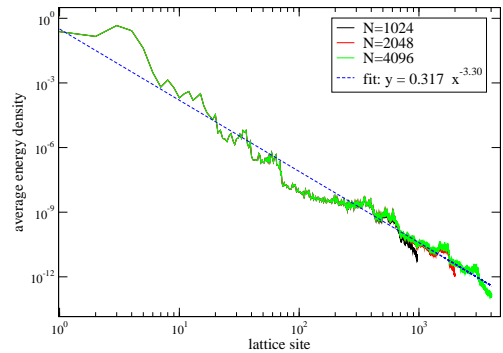}
\end{center}
\caption{The time-averaged energy profile $\overline e_n$ below threshold, 
for $A=0.83322$, $\omega=3.0237$ and different lattice sizes. Other 
parameters like in figs.\ \ref{flx}-\ref{qperio2}.
}
\label{eprof}
\end{figure}

Below threshold energy does not propagate and, as in the DNLS case, only
a few sites close to the driven boundary oscillate. 
As 
shown in fig.\ \ref{qperio2} their motion is quasiperiodic, with
spectra displaying peaks at frequencies 
of the form $m_1\Omega_1+\ldots + m_p\Omega_p$. Each site oscillates 
with a different set of frequencies $\Omega_i$.
To characterize the localization properties of the quasiperiodic state
we computed the time-averaged energy density $\overline e_n$, 
\begin{equation}
\label{localenergy}
e_n \;=\; \frac{\dot u^2_n}{2}+\frac12 
\left[V_{n+1}(u_{n+1}-u_{n})+V_n(u_{n}-u_{n-1})\right] ,
\end{equation}
with $V_n(x)=K_n x^2/2+x^4/4$. 
As seen in fig.\ \ref{eprof}, it displays a 
slow decay along the chain compatible with a power-law, 
$\overline e_n \propto n^{-3.3}$. 
A tail with similar exponent was found also when averaging  over some 
different 
disorder realizations in the nontransmitting  regime.
{Presumably, this is due to
the presence of long-wavelength almost extended modes in the
linear spectrum \cite{MI70}; in DNLS an exponential decay is generally 
seen for the sub-threshold quasiperiodic state.}


The interpretation of the threshold as a transition from quasiperiodicity to 
chaos can be 
seen in the Fourier spectra of 
$\dot u_n$
as an immediate
broadening of the lines 
for $A>A_{\mathrm{th}}$. 
Asymptotically, for the transmitting state 
$\overline e_n$ is found to
reach a given profile as in the ordered
case \cite{khomeriki},  reminiscent of 
stationary heat transport with two thermal baths \cite{rep}. 

In summary, we have shown the existence of well-defined transmission 
thresholds for generic classes of locally time-periodically driven nonlinear 
disordered Hamiltonian chains. An adiabatic threshold for the driver amplitude 
can be defined from the associated NLRM, and 
is nonzero except at a discrete set of 
resonant frequencies. Applying 
the force nonadiabatically 
generates quasiperiodic solutions in the nontransmitting regime, and lowers
the transmission threshold. Beyond the threshold, energy is transmitted 
diffusively through a chaotic state.
These thresholds could be directly observable, {\it e.g.}, 
in optical waveguide arrays \cite{Khom, Lahini}
and for Bose-Einstein condensates in disordered potentials 
\cite{Paul}. 
Although we here focused on two 
particular models, we have found qualitatively similar results also for 
random Klein-Gordon lattices \cite{KAPRL,KA00,MKA06} and the parametrically 
driven 
DNLS model \cite{Hennig}. 
A recent work \cite{Pikovsky} 
also reported a transmission threshold in a model similar to 
eq.\ (\ref{DNLS1}), but with dissipation at both edges. However, while 
ref.\ \cite{Pikovsky} argues that their threshold in average should 
decrease to zero in the limit of an infinite chain, a major conclusion from 
our work is that the threshold generically {\em remains nonzero} also in
the limit $N\rightarrow \infty$ for any typical disorder realization. 
This follows also from the fact that the NLRM generally has a finite slope at 
the origin (except for the zero-measure set of resonant frequencies), 
which was proven in ref.\ \cite{KA00} under basic smoothness assumptions. 
{The existence of a smooth NLRM in the thermodynamic limit 
could possibly also be rigorously proven, at least for the DNLS case, 
as a consequence of the Ruelle-Oseledec theorem.}

This work was initiated within the Advanced Study 
Group 2007 \textit{Localizing energy through nonlinearity, discreteness 
and disorder} at the MPIPKS, Dresden. S.L.\ acknowledges useful discussions 
with S.\ Ruffo. M.J.\ acknowledges support from the 
Swedish Research Council. G.K.\ and S.A.\ thank the Greek GSRT and Egide 
for support through the Platon program.


\end{document}